\documentstyle[12pt,aps,floats,epsf,epsfig]{revtex}

\title{\bf Production of (super)heavy quarkonia and new
Higgs physics at hadron colliders}
\author{
G.A. Kozlov, A.N. Sissakian, J.I. Khubua, G. Arabidze and G. Khoriauli\\
 Joint Institute for Nuclear Research,\\
 141980 Dubna, Moscow Region, Russia\\
e-mail: kozlov@thsun1.jinr.ru\\
and\\
T. Morii\\
 Div. of Sciences for Natural Environment,\\
 Faculty of Human Development,\\
 Kobe University, Kobe, Japan\\
e-mail: morii@kobe-u.ac.jp }

\begin{document}
\maketitle
\begin{abstract}

{\small Based on the two Higgs doublet model, we study the effect of
Higgs-boson exchange on the (super)heavy quarkonium $\bar QQ$,
which induces a strong attractive force between a (super)heavy
quark $Q$ and an antiquark $\bar Q$.  An interesting application
is the decay of (super)heavy quarkonia $\bar QQ$ into a
Higgs boson associated with gauge bosons. The criterion for making
the $\bar QQ$ bound state is studied.  We also show that
non-perturbative effects due to gluonic field fluctuations are
rather small in such a heavy quark sector. Possible enhancement
for productions and decays of $\bar QQ$ bound states made from
the fourth generation quark $Q$ is discussed for $\bar p p$ (at
the Tevatron) and $pp$ (at the LHC) collisions.

PACS 12.38.Aw, 12.38.Lg, 12.40.Qq, 14.80.Dq, 14.80.Gt, 14.80.Er  }
\end{abstract}

\section{Introduction}

A study on quarkonia $T(\bar QQ)$ composed of a (super)heavy quark
$Q$ and an antiquark $\bar Q$ (a possible and interesting candidate of
$Q$ is the up($U$)- and/or down($D$)-quarks in the fourth generation
family) is required in current particle physics for testing the standard
model (SM) and/or searching for signals for physics beyond the SM.
It is nowadays one of the most interesting subjects since the subject
can be studied, with high priority, in the forthcoming experiments
at high energy hadron colliders, i.e. the Tevatron and LHC.
In particular, no theoretical arguments
are seen to rule out the (super)heavy quarks and
the (super)heavy quarkonium states with the
masses around hundred GeV or even a few TeV (see, e.g. [1]).
As the recent example, in one of the extended models, the little
Higgs model [2], there could be an additional
heavy quark $\tilde q$ with a mass of order O(1 TeV) to promote the
quark doublet $q$ to quark triplet under the global SU(3), $
Q=(q,\tilde q)$. By preserving the global symmetry of the
coupling, the one-loop quadratic divergence to the top quark
is removed.
In fact, the SM and its extensions, e.g., the Minimal Supersymmetric
Standard Model (MSSM), do not explain the family (generation)
 structure
of the quark masses.  Each quark has an arbitrary Yukawa coupling and hence
is independent of the family to which it belongs.
It is required to explain the family structure and the
Cabibbo-Kobayashi-Maskawa (CKM) matrix for the quark sector in any
extension of the SM or even in the SM.

It is known from the history of particle physics that the first
signals for $c$- and $b$-quarks in hadronic collisions were leptonic
decays of their $J/\psi (\bar{c} c)$ or $\Upsilon (\bar{b} b)$
bound states.   Can (super)heavy quarks also be first
discovered through the decay of their quark-antiquark bound states
into lepton pairs?  It seems the answer is apparently not transparent,
because one of the main properties of (super)heavy
quarkonia is concerned with the appearance of new decay modes into
weak bosons and even Higgs bosons in the final states.
Once the Higgs boson $H$ is discovered, one needs to measure its
couplings to other particles.  The value of $H$-boson couplings
can be extracted by measuring a variety of Higgs boson productions
and their decay modes.  Thus, it is important to find the $H$-boson
in as many channels as possible, including its production coming from
the decay of (super)heavy quarkonia.
Such a Higgs boson can be looked for in the following new decay modes;
$T(\bar Q Q) \rightarrow HZ,$
$T(\bar Q_{1} Q_{1}) \rightarrow HT(\bar Q_{2} Q_{2}),$
 $T(\bar Q Q) \rightarrow
\gamma H$ and $T(\bar Q Q) \rightarrow ggH, \gamma\gamma H$, where
$T(\bar Q Q)$ carries quantum numbers of $J^{PC}=1^{--}, 0^{-+}$.
Production of heavy quarkonia like $\bar bb$ and $\bar tt$
associated with a Higgs boson emission in the decay of extra gauge bosons
$Z^{\prime}$ has been studied in [3,4].   Here we are interested in
the case in which the new decay modes mentioned above become dominant and
hence the branching ratio for a single heavy quark decay accompanying
the real weak boson emission $Q\rightarrow q W$ ($q$ is a lighter quark),
leaving $q$ as a spectator, is small.  Of course, it is necessary
to examine whether the spectator mode can be dominant or not.

The direct decay of the top quark, $t\rightarrow W b$ [5,6], and
the flavor changing top quark decays, $t\rightarrow c\gamma (c g,~cZ)$
and $t\rightarrow c H$ [7], in the both SM and two-Higgs doublet
model (2HDM), have been studied intensively for the last decade.
The SM predictions of the branching ratios
($BR$) for those decays $t\rightarrow c \gamma$, $t\rightarrow c g$,
$t\rightarrow c Z$ and $t\rightarrow c H$ are significantly small, being
$BR\sim 5\times 10^{-13}$, $4\times 10^{-11}$,  $1.3 \times 10^{-13}$ and
$10^{-14} -  10^{-13}$, respectively.  Obviously, the rare decays are
out of interest here since they are very difficult to be observed
in hadron colliders even at the highest luminosity and thus we neglect
them in this work.
The sensitivity of current experiments at Tevatron Run II or future
experiments at forthcoming LHC closely approaches the rate required
for ruling out the Higgs boson production or discovering it
through the decay of
(super)heavy quarkonia mentioned above.  In this connection, precise
theoretical estimates of the rates are required for an unambiguous
interpretation of experimental upper limits.

One cannot exclude the possibility of the new strong interactions which
primarily control the dynamics of (super)heavy quarks such as
the 4th generation up($U$) and/or down($D$) quarks.
In one of the ``top-color'' models [8] with 1 TeV scale,
there is the following ``top-color'' gauge structure
\begin{eqnarray}
\label{e1}
SU(3)_{4}\times SU(3)_{h}\times SU(3)_{l}\times U(1)_{Y_{4}}
\times U(1)_{Y_{h}}\times
U(1)_{Y_{l}}\times SU(2)_{L}\cr
\rightarrow SU(3)_{QCD}\times U(1)_{EM}\,\,\, ,
\end{eqnarray}
where $SU(3)_{4}\times U(1)_{Y_{4}}$, $SU(3)_{h}\times U(1)_{Y_{h}}$ and
$SU(3)_{l}\times U(1)_{Y_{l}}$ generally coupled to the 4th, 3rd and
first two generations, respectively.   The $U(1)_{Y_{i}}$ are
just rescaled versions of electroweak $U(1)_{Y}$ into the strongly
interacting world.
In this model, below the symmetry-breaking scale $\mu_{SB}$,
the spectrum includes massive ``top-gluons'', which mediate vectorial
color-octet interactions among (super)heavy quarks $Q(=t, U, D)$
\begin{eqnarray}
\label{e2}
-(4\,\pi\,\kappa/\mu_{SB}^2)\left (\bar Q\gamma_{\mu}\frac{\lambda^a}{2}\,
 Q\right )^{2}\, .
\end{eqnarray}
If the coupling $\kappa$ lies above some critical value $\kappa_{crit}$,
the heavy quark condensate $\langle \bar Q Q\rangle$ can be formed.
The strong ``top-color'' dynamics can bind $\bar Q$ and $Q$ into a
set of ``heavy-pions'' $(\bar Q Q)$.
The criterion for existence of a (super)heavy quarkonium is that the binding
energy $\epsilon_{B}$ should be larger at least than the total decay width
$\Gamma_{tot}$ of its quarkonium, namely $c=(\Gamma_{tot}/\epsilon_{B}) < 1$.
Since such a (super)heavy quark-antiquark bound state is considered to be a
non-relativistic system, the quark potential model should be applicable to the
analysis.  Then as pointed out in [9-11], one cannot exclude the possibility
of the Higgs-boson interaction which dominates significantly over the
one-gluon exchange $\sim -(4/3)\alpha_{s}(m_{Q})/r$ for the (super)heavy
quarkonium, where $\alpha_{s}$ is the strong coupling constant depending
on the quark mass $m_{Q}$ and $r$ is a distance between a quark and an
antiquark.  We show in Sec.2 that the strong binding force due to the
Higgs-boson exchange gives rise to a necessary condition
$c < 1$ for enabling the heavy quarkonium to exist and leading to
observation of its resonance.

On the other hand, in the physics of interplay among quarks,
it is well known that the exact QCD vacuum should contain the
fluctuations of gluonic fields at large scales [12].
Those non-perturbative fluctuations cause the distortion
of interactions between quarks and antiquarks. We consider (super)heavy
quarks as external objects allocated in the gluonic vacuum.
Here, we study those non-perturbative fluctuation effects
of the gluon field on the decay of a quarkonium $T(\bar U U)$ into the
Higgs- and $Z$-bosons. Our result is based on the well-known
statement (see, e.g., [12]) that the non-perturbative effect on dynamics
of heavy quark systems is expressed in terms of vacuum expectation values
of the local operators constructed from gluonic field operators.
The leading effect is proportional to a matrix element
of the form ${\langle {{G^a}_{\mu\nu}}^2(0)\rangle}_{0}$, where
$G_{\mu\nu}^{a}(x)$ is the standard gluonic field strength tensor
with color indices $a=1,2,\cdots ,8$.
The lowest level of the (super)heavy quarkonium is determined
by the color-singlet Yukawa-type attractive force
mediated by the ``light'' scalar $\chi$-boson. Here,
we are interested in the corrections due to
non-perturbative gluonic fluctuations in the exact QCD vacuum.
In this paper, we show that in a (super)heavy quarkonium decay such as
$T(\bar U U)\rightarrow h Z$ ($h$ means a lightest CP-even Higgs-boson
in 2HDM), the non-perturbative fluctuation
effect of the gluonic field can be calculated, to some extent,
without detailed knowledge of the vacuum structure and furthermore,
it gives a negligible result.

The outline of this work is organized as follows.  In Sec. 2, we
discuss an effective potential mediated by a Higgs boson.
An effective model for the lower-energy theorem will be discussed
in Sec. 3.  Finally, in Sec. 4, we give our conclusion and discussion.

\section{Effective potential via Higgs boson exchange}

First of all, one should notice that the total cross-section for
the process $pp(\bar p p)\rightarrow \bar Q Q$ is strongly related to the
underlying subprocesses. For example, at $\sqrt s=$14 TeV,
the cross section for various processes of $Q$ and $\bar Q$ pair
productions with $Q=U$ and/or $D$ in the mass region
$m_{U}\simeq m_{D}\simeq m_{4}=0.2-0.6$ TeV ($m_{4}$ is the mass
of the fourth generation quark), becomes $\sigma\simeq 10^{-1}-10^{-2}$ (pb)
and $\sigma\simeq 1-10^{-2}$ (pb) for $\gamma/Z$ production and $W$
production, respectively, in the $q$ and $\bar q$ annihilation channels,
and $\sigma\simeq 10-10^{-1}$ (pb) for the two-gluon fusion channel.

Let us consider a production (in $pp$ or $\bar p p$) of (super)heavy
quarkonium $T(\bar Q Q)$ followed by the decay process
$T(\bar Q Q)\rightarrow h Z$ with $Q=t, U, D$, being assumed to be
the dominant decay process of $T(\bar QQ)$.  Then, it is supposed
that this dominant mechanism at high transverse momentum involves, e.g.
the production of a gluon that produces a color-octet $\bar Q$ and $Q$ pair
which then fragments into a color-singlet bound state $T(\bar Q Q)$
by emitting two or more soft gluons.
This $T(\bar Q Q)$-state should be transversely polarized at high transverse
momentum, since it is emanated from a gluon which has only
transverse polarization states.

In the lowest bound state $\bar QQ$, the quark $Q$ and the antiquark
$\bar Q$ are assumed to be located at a distance
\begin{eqnarray}
\label{e3}
r\sim [m_{Q}\, \lambda (m_{Q})]^{-1}
\end{eqnarray}
which is small compared to the scale of strong interactions.
$\lambda (m_{Q})$ is a strength of the interaction between a quark
and an antiquark.  The wavefunction of the lowest bound state is
proportional to $\exp (-\mu\,r)$ with $\mu\sim m_{Q}\, \lambda (m_{Q})$.
Note that for $r_{0}=\mu^{-1}$ being smaller than in the typical
size of fluctuations, our approach becomes applicable
since the potential dumps exponentially for a distances $r\gg \mu^{-1}$.

Let us consider a simple model (Model I) where the dominant effective
potential for a $\bar Q$ and $Q$ system at a small distance looks like
[10,11]
\begin{eqnarray}
\label{e4}
V_{eff}(r)\sim -\frac{C_{F}}{r}\, \alpha_{s} (m_{Q})-
\frac{\lambda(m_{Q},\xi_{
\chi\,Q})}{r}\,\exp (-m_{\chi}\, r)\, ,
\end{eqnarray}
with
\begin{eqnarray}
\label{e5}
\lambda(m_{Q},\xi_{\chi Q})=\frac{m_{Q}^2}{4\,\pi\,v^2}\,\xi^{2}_{\chi\,Q}\, ,
\end{eqnarray}
and $\xi_{\chi\,Q}$ reflects the model ``flavor'' in the strength
of the interaction between the scalar $\chi$-boson and a heavy quark
$Q$ ($\xi_{\chi\,Q}=1$ in the minimal SM, otherwise $\xi_{\chi\,Q}>1$).
$v$ is the vacuum expectation value of Higgs boson, $v$= 246 GeV (in
 2HDM, $v=\sqrt{v_1^2+v_2^2}$, $v_{1}$ and $v_{2}$
are two neutral Higgs field vacuum expectation values) and $C_F$
is the color
factor, $C_{F}=4/3$ for color $SU(3)$ group. In Fig.1, we show the ratio
of the combined coupling $\alpha_{comb}=(4/3)\alpha_{s}(m_{Q})
+\lambda(m_{Q},\xi_{\chi\,Q})\cdot\exp (-m_{\chi}\, r)$ (see (\ref{e4}))
to the pure QCD coupling $\bar\alpha_{s}=(4/3)\alpha_{s}$, as a
function of a heavy quark mass $m_{Q}$ for different values of
$\xi_{\chi Q}$ and $m_{\chi}$.  The second term in $\alpha_{comb}$ is
appropriate for $r\sim [m_{Q}\,\lambda (m_{Q})]^{-1}$.
The ratio becomes somewhat bigger for smaller Higgs-boson masses.
\begin{center}
\begin{figure}[h]
\resizebox{15cm}{!}{\includegraphics{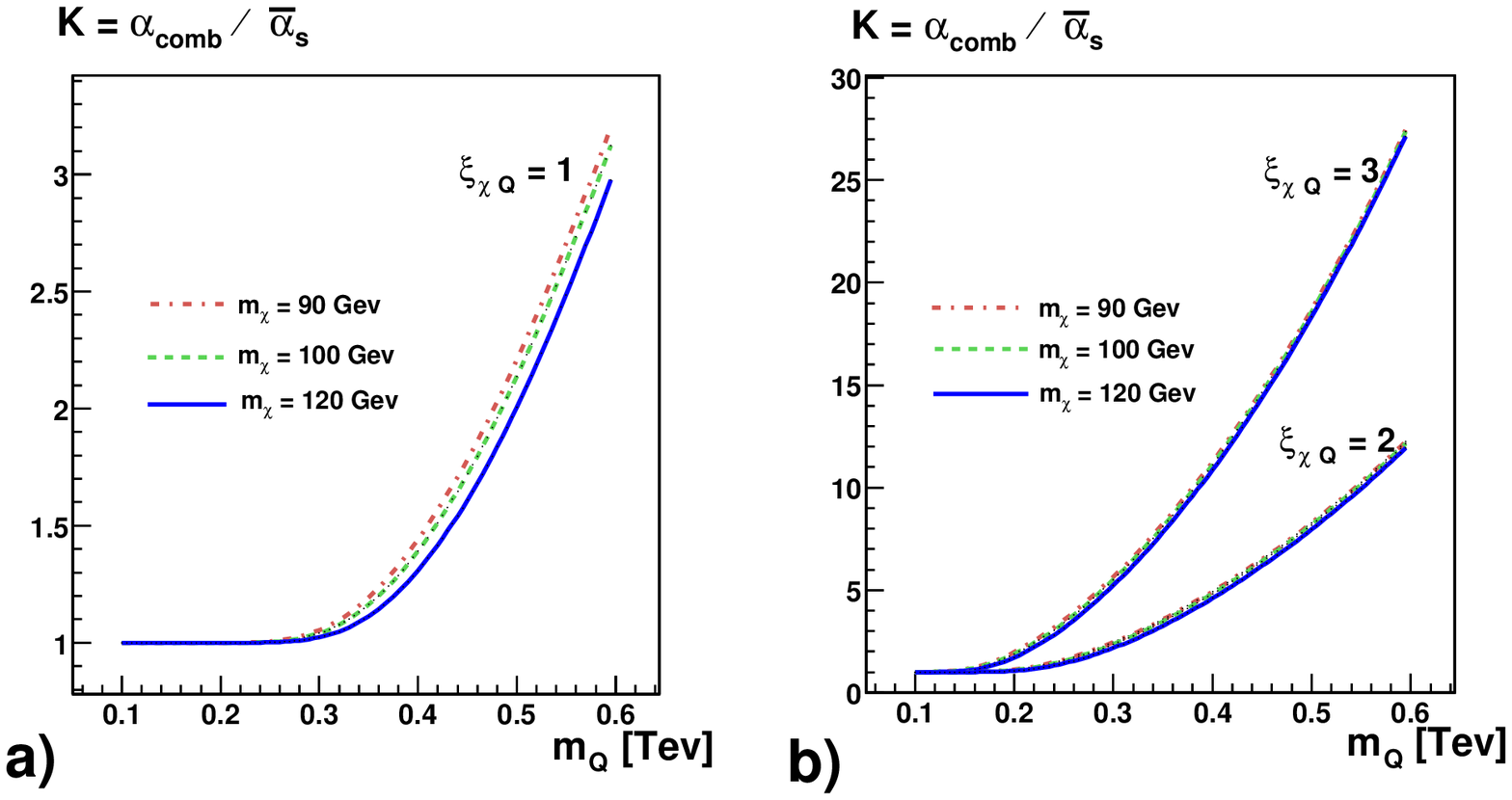}}

Fig. 1 Ratio of the combined coupling $\alpha_{comb}$ to the pure QCD
coupling $\bar\alpha_{s}$ as a function of a heavy quark mass
for (a) $\xi_{\chi\,Q} =1$ and (b) $\xi_{\chi\,Q} =2,3$. The curves are
presented for $\chi$-boson masses $m_{\chi} = 90, 100$ and $120$ GeV.

\end{figure}
\end{center}

Because of our demand,
$\lambda(m_{Q},\xi_{Q\,\chi})> C_{F}\,\alpha_{s}(m_{Q})$,
for a relevance of the $\chi$-boson interaction, the lower
bound on $m_{Q}$ is given as
\begin{eqnarray}
\label{e6}
m_{Q}>\frac{v}{\xi_{\chi\, Q}}\,(4\,\pi\,C_{F}\,\alpha_{s})^{1/2}\, ,
\end{eqnarray}
which leads to $m_{Q}\geq m_{t}$ even if $\xi_{\chi\, Q}$= 2
(see Fig.2).\\


\begin{center}
\begin{figure}[h]

\resizebox{12cm}{!}{\includegraphics{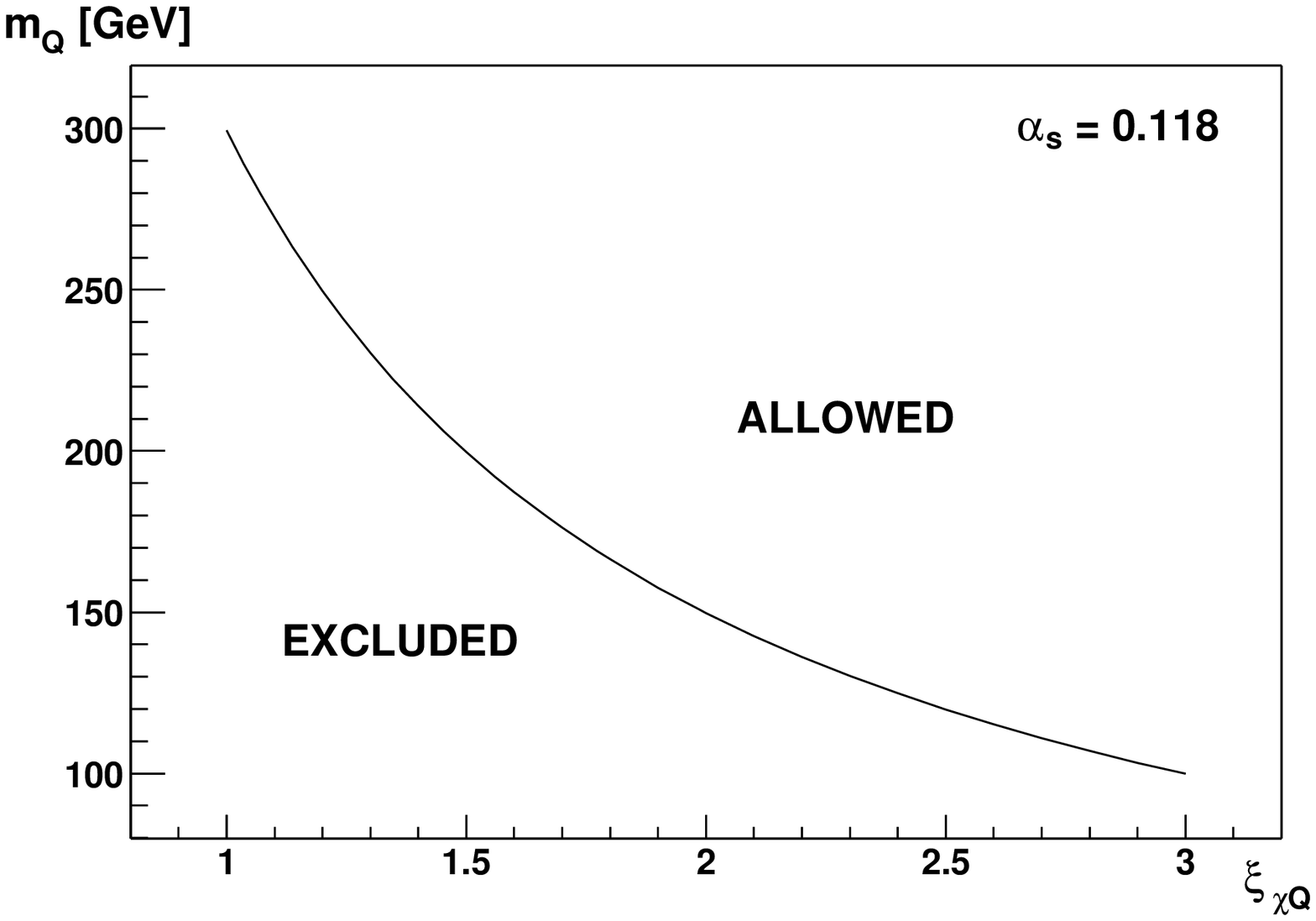}}

Fig.2  Lower bound on (super)heavy quark masses as a function of
$\xi_{\chi\,Q}$.\\

\end{figure}
\end{center}


The requirement of the positivity of the variational parameter
\begin{eqnarray}
\label{e7}
\mu\simeq \frac{\lambda\,m_{Q}}{2}\,\frac{ (\lambda\,m_{Q})^{2}-m_{\chi}^{2}}
{(\lambda\,m_{Q})^{2}+2\,m_{\chi}^{2}}
\end{eqnarray}
entered in both the bound state wavefunction
$\Psi(r)=2\,\mu^{-3/2}\,\exp(-\mu\,r)$ and the binding energy
(see, for details, [11]),
\begin{eqnarray}
\label{e8}
\epsilon_{B}=2\,\lambda\,\frac{\mu^3 (2\,\mu-m_{\chi})}{(2\,\mu+m_{\chi})^{3}}
\end{eqnarray}
leads to an upper limit on $m_{\chi}$ (see Fig.3).\\

\begin{center}
\begin{figure}[h]
\resizebox{12cm}{!}{\includegraphics{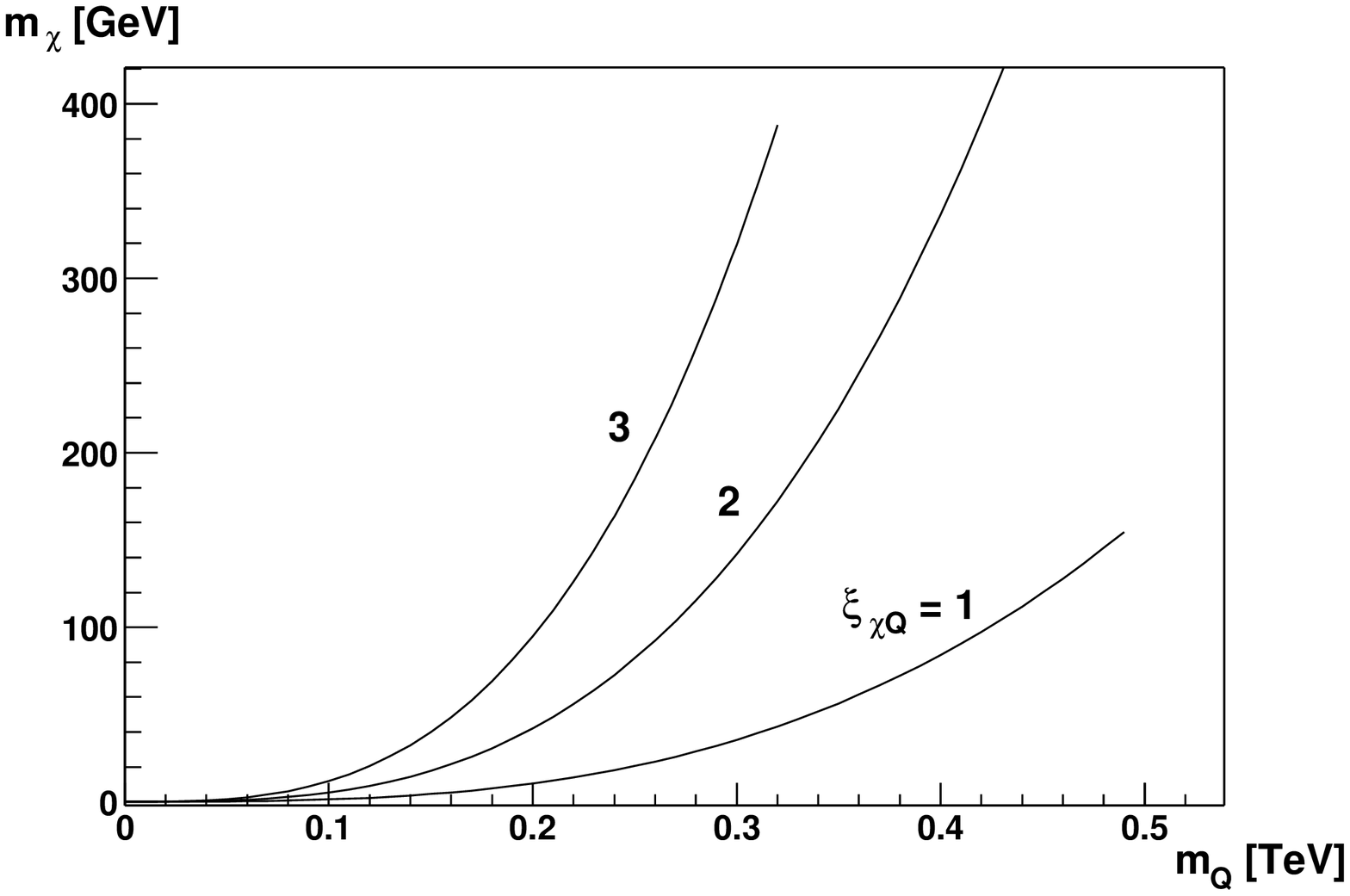}}

Fig.3 Upper limit on the scalar  $\chi$-boson mass as a function
of $m_{Q}$ for different values of $\xi_{\chi\, Q}$. The regions above the
corresponding curves are excluded.

\end{figure}
\end{center}


In the Model I, the ratio $c=(\Gamma_{tot}/\epsilon_{B})<1$ could be
guaranteed for $T(\bar U U)$ quarkonia to be formed by a strong attractive
force via scalar Higgs-boson exchange with a sufficiently
``hard'' Yukawa coupling $\lambda(m_{Q},\xi_{\chi\, Q})$.  The total
decay width $\Gamma_{tot}$ is given by a sum of two terms
\begin{eqnarray}
\label{e9}
\Gamma_{tot}=\Gamma_{T}+\Gamma_{U}\, ,
\end{eqnarray}
where the width $\Gamma_{T}$ is defined by the following decay channels
$$T(\bar U U)\rightarrow h Z, \gamma Z, \gamma h, W^{+}W^{-}, \bar b b,
\bar t t, \tau^{+}\tau^{-}, \mu^{+}\mu^{-}, ggg\, ,$$
and the single quark decay width $\Gamma_{U}$ is a sum of the following
contributions: $U\rightarrow D W^{+}, b W^{+}, b H^{+}$ ($H^{+}$ is the
charged Higgs-boson in 2HDM).  In the decays of $T(\bar UU)$ presented
above, the radiative channel is suppressed by the coupling constant $\alpha$,
the $ggg$ channel gives the small contribution due to
$\alpha_{s}^{3}$ factor, and furthermore, the productions of the pairs
 of quarks and
antiquarks or leptons and antileptons are also small because they
follow via two-loop diagrams, where the amplitude is very small because
of the presence of $\alpha$ (intermediate photons) or the Fermi constant
$G_{F}$ (virtual $W^{\pm}$-bosons).

We do not consider the contributions from the decay
$U\rightarrow  b H^{+}$ because it is expected to have a rather
small probability to be observed as expected from the following
consideration: the Higgs-boson mass sum rule
\begin{eqnarray}
\label{e10}
m_{H^{+}}^{2}\simeq (m_{A}^{2}+m_{W}^{2})\,(1+\delta)
\end{eqnarray}
with one-loop correction $\delta< 10 \%$ [13] does not allow the production
of charged Higgs-bosons in the decay of top quark and, perhaps, also of
$U$ quark due to the kinematical reason in the decoupling limit,
$(m_{W}^{2}/m_{A}^{2})\ll$ 1 ($m_{A}$ is the CP-odd Higgs-boson in 2HDM).
In addition, the experimental data at the Tevatron do not yet clarify the
status of $t\rightarrow H^{+} b$ decay.  The CDF results in direct search
of a $\tau$-lepton emission from top quark decays give an upper limit on
the branching ratio $BR(t\rightarrow H^{+} b)\sim 0.5-0.6$ at $95 \%$ C.L.
in the range $60~ GeV < m_{H^{+}} < 160~ GeV$, assuming
 $BR(H^{+}\rightarrow\tau\nu_{\tau}) =1$ [14].  Furthermore,
the D0 Collaboration
excludes $BR(t\rightarrow H^{+} b)> 0.36$ at $95 \%$ C.L.
in the region $0.3 <\tan\beta <150$ and $m_{H^{+}} <160~GeV$ [15].
Assuming $\sum_{X=H^{+},W^{+}}BR(t\rightarrow b X) =1$,
the decay width of the $H^{+}$-boson channel in $U$- and $t$-quark
decays is small and hence this channel is out of interest.

Therefore, for the case of $U$ quarks,
one can expect that the $\Gamma_{tot}$ is given by
\begin{eqnarray}
\label{e11}
\Gamma_{tot}=\Gamma_{T}(T(\bar U U) \rightarrow h Z, W^{+}W^{-})+
\Gamma_{U}(U\rightarrow D W^{+}, b W^{+})\, .
\end{eqnarray}
The main contribution to $\Gamma_{T}$ arises from the channel
$T(\bar U U)\rightarrow h Z$,
whose decay width for $T(\bar UU)(1^{--})$ is given by the
following expression (see, e.g., [11])
\begin{eqnarray}
\label{e12}
\Gamma(T(\bar U U)\rightarrow h Z)=
\frac{\lambda^{3}\,m_{U}}{16}\,\eta^{2}_{h\,U}\,
\left [\lambda\,\alpha_{z}\,v_{U}^{2}+\frac{1}{2}\,\alpha_{W}^{2}\left
(\frac{m_{U}}{m_{W}}\right )^{4}\right ]\, f^{3}\,\Phi\, ,
\end{eqnarray}
with
$$\alpha_{W}=\alpha/s_{W}^{2}\, ,\alpha_{z}=\alpha_{W}/c_{W}^{2}\,\,\,
 (s_{W}\equiv\sin\theta_{W},
c_{W}\equiv\cos\theta_{W}), $$
$$v_{U}=\left (1-\frac{8}{3}\,s^{2}_{W}\right )\,,~
\Phi=\left (1-\frac{4\,m_{h}^{2}}
{M_{T}^{2}}\right )^{1/2}\, ,$$
$$f\equiv f(\lambda\, ,m_{U},m_{\chi})=\frac{(\lambda\,m_{U})^{2}-m_{\chi}^{2}}
{(\lambda\,m_{U})^{2}+2\,m_{\chi}^{2}}\,\,\, ,M_{T}\simeq 2\,~m_{U}\, .$$
$m_{U}$ is the mass of an up($U$)-quark of the fourth generation.
We suppose that the couplings of the Higgs-boson $h$ and the $U$-quark
have the same form as those for the couplings of the Higgs-boson $h$ and
the top-quark (see the review [16])
\begin{eqnarray}
\label{e13}
\eta_{h\,U}\simeq 1+\frac{m_{Z}^2}{m_{A}^{2}}\,s_{2\,\beta}\,c_{2\,\beta}\,
\tan^{-1}{\beta}\,  ,
\end{eqnarray}
where $s_{2\,\beta}(c_{2\,\beta})\equiv\sin 2\,\beta (\cos 2\,\beta)$
and $\tan\beta$ is the standard ratio between two vacuum expectation
values for two Higgs doublets in 2HDM.

Using the standard formulae [5,6], the single $U$-quark decays are given
for $U\rightarrow D\,W^{+}$ channel by
%
\begin{eqnarray}
\label{e14}
\Gamma_{U}(U\rightarrow D W^{+})=
\frac{G_{F}\,m_{U}^{3}}{8\,\sqrt 2\,\pi}\,{\vert V_{UD}\vert}^{2}
\left [\left (1-\frac{m_{D}^{2}}{m_{U}^{2}}-
\frac{m_{W}^{2}}{m_{U}^{2}}\right )^{2}-
4\,\frac{m_{D}^{2}\,m_{W}^{2}}{m_{U}^{4}}\right ]^{1/2}\cr
\times\left [\left (1-\frac{m_{D}^{2}}{m_{U}^{2}}\right )^{2}+
\left (1+\frac{m_{D}^{2}}{m_{U}^{2}}\right )\cdot\frac{m_{W}^{2}}{m_{U}^{2}}-
2\,\frac{m_{W}^{4}}{m_{U}^{4}}\right ]\, ,
\end{eqnarray}
and for $U\rightarrow b\,W^{+}$ decay by
\begin{eqnarray}
\label{e15}
\Gamma_{U}(U\rightarrow b W^{+})=
\Gamma_{0}(U\rightarrow b W^{+})\,(1-\Delta)\, ,
\end{eqnarray}
where
\begin{eqnarray}
\label{e16}
\Gamma_{0}(U\rightarrow b W^{+})=
\frac{G_{F}\,m_{U}^{3}}{8\,\sqrt 2\,\pi}\,{\vert V_{Ub}\vert}^{2}
\beta_{W}^{4}\,(3-2\,\beta_{W}^{2})
\end{eqnarray}
with $\beta_{W}=\sqrt{1-m_{W}^{2}/m_{U}^{2}}$. $V_{UD}$ and $V_{Ub}$
are generalized CKM matrix elements and the $O(\alpha_{s})$ QCD
correction [6] in (\ref{e15}) is
\begin{eqnarray}
\label{e17}
\Delta =\frac{C_{F}\,\alpha_{s}}{2\,\pi}\left (\frac{2\,\pi^2}{3}
-\frac{5}{2}\right )\,.
\end{eqnarray}
As an example, for a typical set of parameters:
 $m_{U}$= 400 GeV, $m_{D}$= 300 GeV,
$m_{\chi}\simeq m_{h}$= 100 GeV, $\vert V_{UD}\vert\simeq 1$,
$\xi_{\chi\,U}=2$ ($\eta_{h\,U}\simeq 1$)
 we give the numerical results for those decay widths
\begin{eqnarray}
\label{e18}
\Gamma(T(\bar U U)\rightarrow h Z)&\simeq& 2.2\,\eta_{h\,U}^{2}~{\rm GeV},
~~~
 \nonumber \\
%
%
\Gamma(U \rightarrow D W^{+} )&\simeq& 1.21~{\rm GeV}\,,\,\,\,
\Gamma(U \rightarrow b W^{+} )\simeq 0.0021~{\rm GeV}\, ,
\end{eqnarray}
where $c=\Gamma_{tot}/\epsilon_{B}=0.27$ at the calculated value
$\epsilon_{B}=12.65~{\rm GeV}$.

The contribution of the decay $U \rightarrow b W^{+}$ is negligible
due to rather small value of $\vert V_{Ub}\vert\sim 10^{-2}$.
We have checked the possibility of the existence
of bound states $T(\bar Q Q)$ composed of $U$ (and $D$)-quarks,
starting at the lowest value of $\xi_{\chi Q} \geq 1$, where the
increasing $\xi_{\chi Q}$ gives rise to an
effect on $c^{-1} > 1$ (see Fig.4).

\begin{center}
\begin{figure}[h]
\resizebox{12cm}{!}{\includegraphics{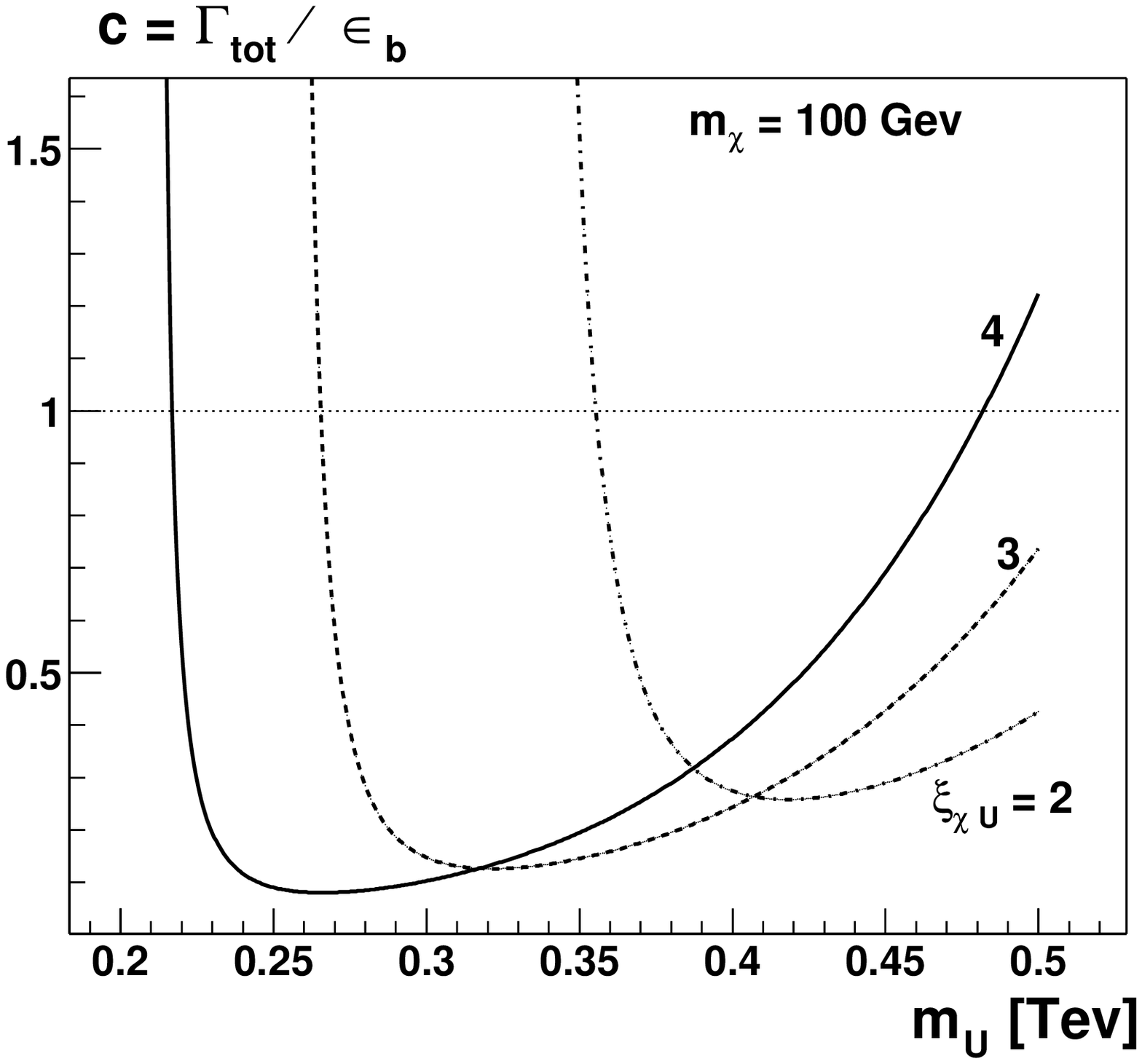}}

Fig.4 Ratio $c=\Gamma_{tot}/\epsilon_{B}$ being responsible for
occurrence of the bound
state $T(\bar U U)$ as a function of the quark mass $m_{U}$ at different
values of $\xi_{\chi U}$ for $m_{\chi}= 100$ GeV.
The region below the value $c=1$ is allowed.\\

\end{figure}
\end{center}
 The expected
event topology of the decays $U\rightarrow b W^{+}$ and $U\rightarrow D W^{+}$
are similar to that of $t\rightarrow b W^{+}$.  However, for the down-type
heavy quark $D$, the dominant decay mode could be $D\rightarrow t W^{-}$
and thus, in the case of $\bar D D$-pair production, the final state can
consist of two pairs of leptons and neutrinos (originated from
decays of the $t$ quark and $W$ boson) with different flavors, in
general. Now, by taking, as an example, the following parameters:
$m_{D}$= 400 GeV, $m_{\chi}\simeq m_{h}$= 100 GeV, $\xi_{\chi\,D}=2$,
$\vert V_{Dt}\vert\simeq $ 0.012, we can obtain the following
decay widths
\begin{equation}
\label{e19}
\Gamma(T(\bar D D)\rightarrow h Z)\simeq 2.25\,
\eta_{h\,D}^{2}~{\rm GeV}\,,~~
\Gamma(D \rightarrow t W )\simeq 1.47~{\rm MeV}\, .
\end{equation}
One can see that the decay width $\Gamma(T(\bar D D)\rightarrow h Z)$
can be enhanced by the Yukawa-flavor factor
(within 2HDM) $\eta_{h\,D}=-\sin\alpha/\cos\beta$ at large
values of $\tan\beta$.

For comparison, let us consider an instructive example, i.e.
the decay of $T(\bar UU)(0^{-+})$ states to $h$- and $Z$-boson,
where the decay width is given by the following expression
(see, e.g., [11])
\begin{equation}
\label{e20}
\Gamma(T(\bar U U)\rightarrow h Z)=\frac{3\,\lambda^{3}\,m_{U}}{32}\,
\eta^{2}_{h\,U}\,
\alpha_{z}^{2}\,\left  (\frac{m_{U}}{m_{Z}}\right )^{4}\, f^{3}\,\Phi\,.
\end{equation}
The numerical estimation gives
$\Gamma(T(\bar U U)\rightarrow h Z)\simeq 6.80\,\eta_{h\,U}^{2}$ GeV,
which is roughly 3 times larger than that in the case of
$T(\bar UU)(1^{--})\rightarrow h Z$ decay mode given in (\ref{e18})
and hence leads to larger production rate of $T(\bar UU)(0^{-+})$
than $T(\bar UU)(1^{--})$.
This simple example confirms our belief that the most promising
candidate of the (super)heavy quarkonium which could be searched at the LHC
should be the pseudoscalar state $T(\bar UU)(0^{-+})$.\\
It is interesting to estimate the effect of the scalar $\chi$-boson
exchange on the ($\bar Q Q$) production cross-section
 at different $m_{\chi}$ as a function of $m_{Q}$.
%
%

If the $Q$-quark is relatively long-lived, the peak of the
cross-section at a $\bar QQ$ resonance due to the one-gluon
exchange is given by
\begin{eqnarray}
\label{e21}
\sigma_{c}\sim\alpha_{s}^{3}\,\left (\frac{m_{Q}}{\Gamma_{Q}}\right )
\,\frac{1}{s}
\end{eqnarray}
for a given center-of-mass energy $s$. The cross-section
(\ref{e21}) has a strong sensitivity to $\alpha_{s}$ and
decreases sharply with increasing $m_{Q}$ because of the
rapid grougth of $\Gamma_{Q}$ (see Eqs. (\ref{e14}) and
(\ref{e15})).
To leading order,
the scalar $\chi$-boson exchange effect is taken
into account simply by making the replacement
\begin{eqnarray}
\label{e22}
\alpha_{s}\rightarrow\alpha_{s}+\tilde\alpha
(m_{Q},\,m_{\chi},\,\xi_{\chi\,Q})\, ,
\end{eqnarray}
where
\begin{eqnarray}
\label{e23}
\tilde\alpha (m_{Q},\xi_{\chi\,Q})=
\frac{3\,m_{Q}^2}{16\,\pi\,v^2}\,
\xi_{\chi\,Q}^{2}\exp(-\epsilon)
\end{eqnarray}
with $\epsilon=m_{\chi}/(m_{Q}\,\lambda)\ll 1$.  Here, we see a
simple increase of the coupling strength between $\bar Q$ and $Q$.
We find a relative enhancement of the effective cross-section
$\sigma_{eff}$
\begin{eqnarray}
\label{e24}
\sigma_{eff}=\sigma_{c}\,\left (1+3\,\tilde\alpha\,\alpha_{s}^{-1}+
3\,\tilde\alpha^{2}\,\alpha_{s}^{-2}+
\tilde\alpha^{3}\,\alpha_{s}^{-3}\right )
\end{eqnarray}
due to the $\chi$-boson exchange effect. In Fig.5 we show the ratio of
 the effective cross-section (\ref{e24}) to the
cross-section (\ref{e21}) as a function of $m_Q$.
The curve in Fig. 6 is appropriate for a massless $\chi$- boson.\\

\begin{center}
\begin{figure}[h]
\resizebox{12cm}{!}{\includegraphics{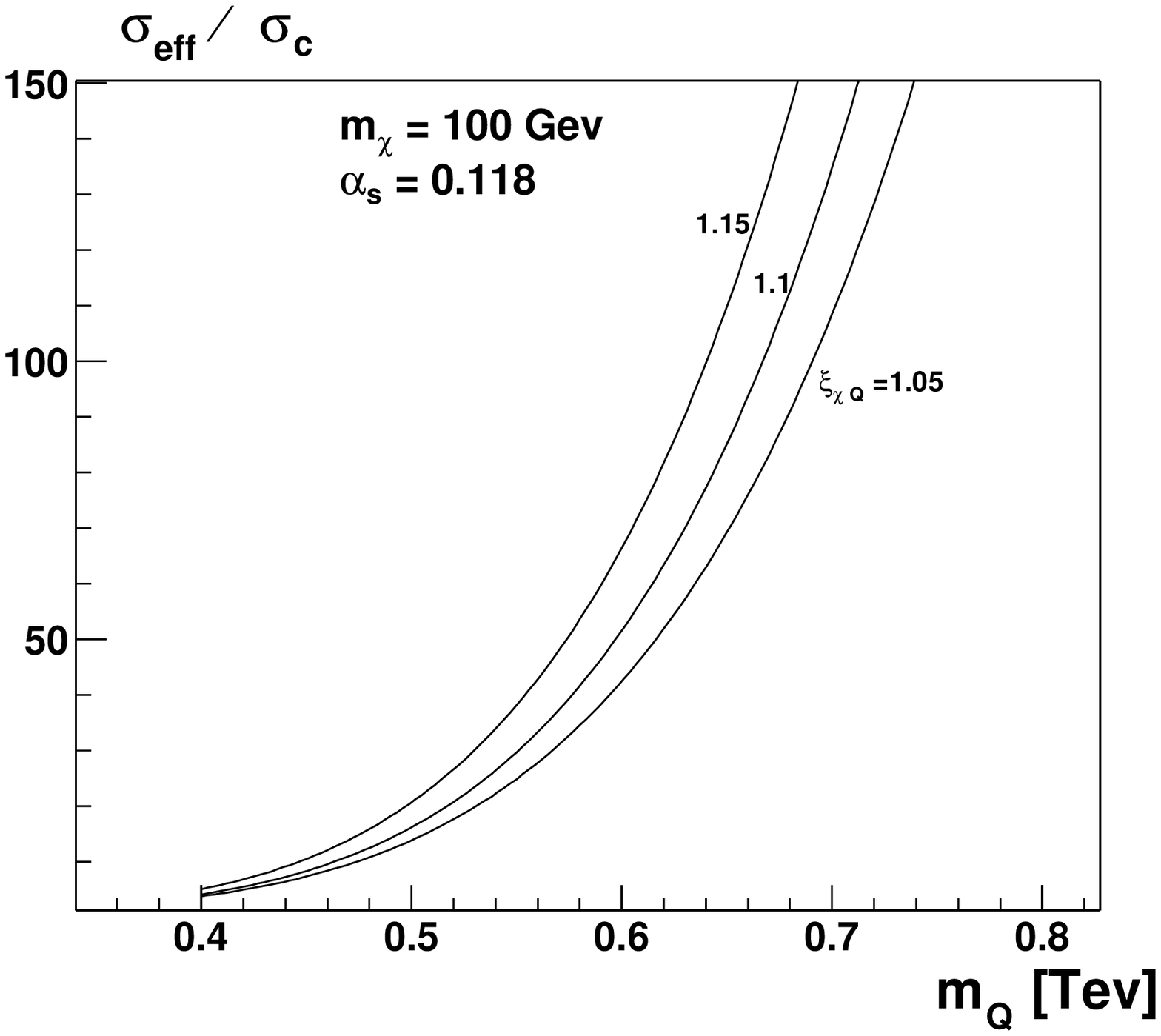}}

Fig.5  Ratio of the effective cross-section $\sigma_{eff}$ (\ref{e24})
and the cross-section $\sigma_{c}$ at the 1S resonance $\bar Q Q$
as a function of $m_{Q}$.  The curves are for
$\xi_{\chi\,Q}$= 1.05, 1.10, 1.15 and
$m_{\chi}$ = 100 GeV at $\alpha_{s}$= 0.118.

\end{figure}
\end{center}


\begin{center}
\begin{figure}[h]
\resizebox{12cm}{!}{\includegraphics{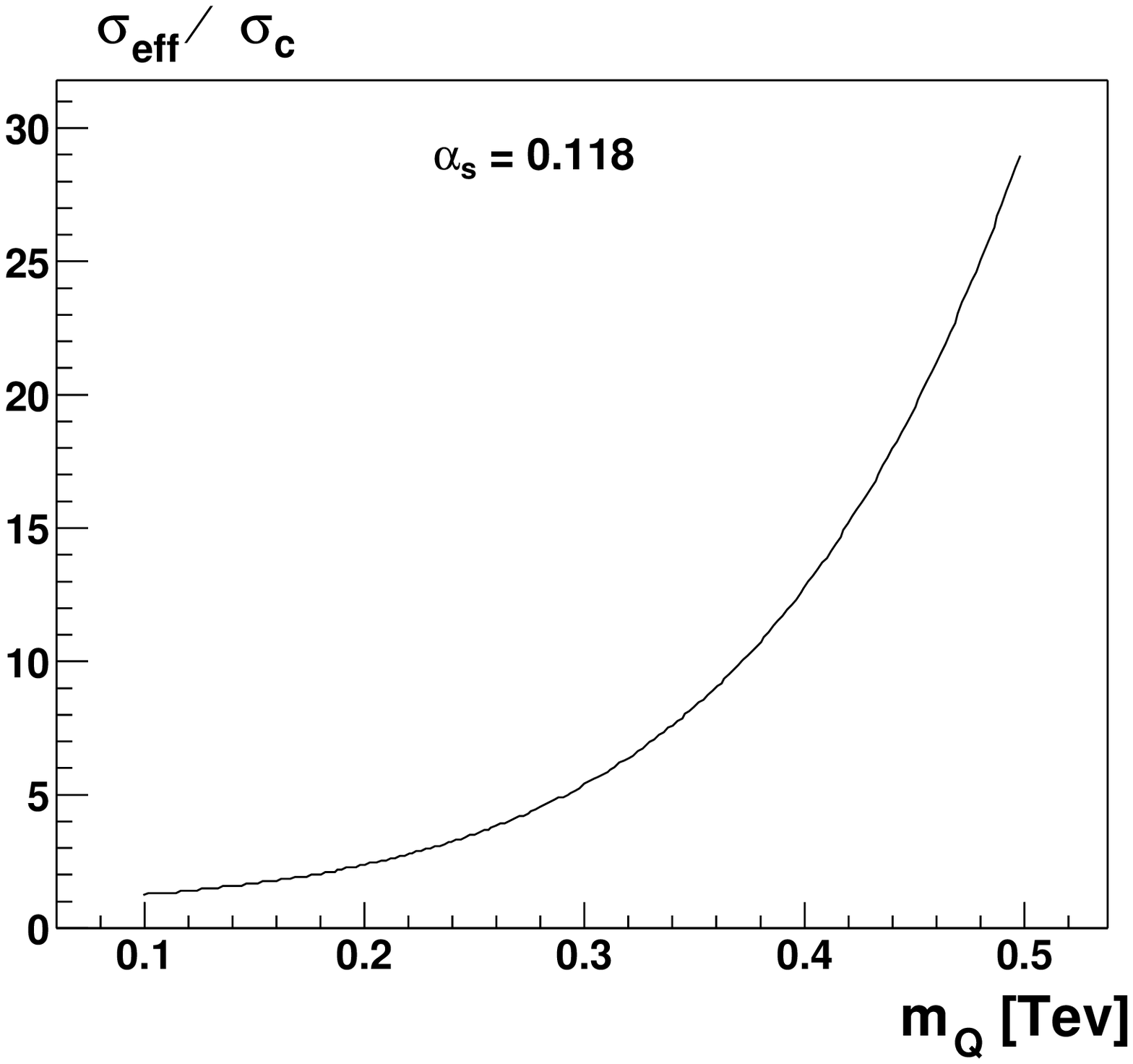}}

Fig.6  Ratio of the effective cross-section $\sigma_{eff}$ (\ref{e24})
and the cross-section $\sigma_{c}$ at the 1S resonance $\bar Q Q$
as a function of $m_{Q}$.  The curve is for the massless
$\chi$-boson.

\end{figure}
\end{center}

In the end of this Section we give briefly the results of
the decays  $\Phi\rightarrow T(\bar Q Q)+\gamma$
where a Higgs-boson $\Phi$ ($\Phi= h$ or $\Phi= H$) with 4-momentum
$q_{\mu}$ and the mass $m_{\Phi}$
decays in the heavy vector ($1^{--}$)
quarkonium $T(\bar Q Q)$ ($Q=b$-quarks,
$t-$quarks, ...), carrying the momentum $P_{\mu}=2\,p_{\mu}$
 $(P^{2}=m_{T}^{2})$, and a photon with the momentum squared $k^{2}=0$.

The amplitude of the transition $\Phi\rightarrow T\,\gamma$ is
[17]
\begin{eqnarray}
\label{e25}
A(\Phi\rightarrow T\,\gamma)=-2\,\sqrt{4\,\pi\,\alpha}\,g_{V}\,e_{Q}
\frac{1}{v}\,
\eta_{\Phi\,Q}\,\frac{1}
{1-(m_{\Phi}/m_{Q})^{2}}\,\epsilon^{\alpha}\,k^{\beta}\,
(P_{\alpha}\,\phi_{\beta}-P_{\beta}\,\phi_{\alpha})\, ,
\end{eqnarray}
where $e_{Q}$ is the charge of the quark $Q$,
$\phi_{\mu}$ being the polarization vector of $T(\bar Q Q)$ and
$g_{V}$ is defined in the standard manner
$$\langle T(\bar Q Q)\vert\bar Q\,\gamma_{\mu}\,Q\vert\, 0\,\rangle
=m_{T}^{2}\,g_{V}\,\phi_{\mu} $$
and can be estimated from the leptonic decay width $T(\bar Q
Q)\rightarrow\bar l l$ ($l=e\, ,\mu\, ,\tau)$:
$$\Gamma (T(\bar Q Q)\rightarrow\bar l l)
=\frac{4}{3}\,\pi\,(\alpha\,e_{Q}\,g_{V})^{2}\,m_{T}\, . $$
At an arbitrary large values of $y= (m_{\Phi}/2\,m_{Q})^{2}>$
1 there are the corrections to the amplitude (\ref{e25}) due to
the one-loop $O(\alpha_{s})$ gluon contributions
$$1-F(y)\,C_{F}\,\alpha_{s}(m_{\Phi}^2-4\,m_{Q}^2)\, ,$$
where the function $F>0$ was calculated in Ref. [17].
For the decay processes $h\rightarrow \Upsilon(\bar b b)\,\gamma$,
 $H\rightarrow \Upsilon(\bar b b)\,\gamma$
and $H\rightarrow T(\bar t t)\,\gamma$ we are considered here as the
promising channels in a wide range of the Higgs-boson mass we
expect the relative decay width compared to final state quarks
$\bar Q Q$, which is given by [17]
\begin{eqnarray}
\label{e26}
f\equiv\frac{BR(\Phi\rightarrow T(\bar Q Q)\,\gamma)}
{BR(\Phi\rightarrow \bar Q Q)}=
\frac{\Gamma(\Phi\rightarrow T(\bar Q Q)\,\gamma)}
{\Gamma(\Phi\rightarrow \bar Q Q)}=  \cr
 \frac{64\,\pi\,\alpha\,e_{Q}^{2}}{3}\,
g_{V}^{2}\,\left
(\frac{m_{Q}}{m_{\Phi}}\right )^{2}\,K^{2}(y ,\alpha_{s})\,
\left [1-\left (\frac{m_{T}}{m_{\Phi}}\right )^{2}\right
]^{-1/2}\, ,
\end{eqnarray}
where
$$K(y ,\alpha_{s})\simeq 1-\frac{\alpha_{s}(m_{\Phi}^{2}-m_{T}^{2})}
{\pi}\,C_{F}\,\ln{2}\,\ln (4y) $$
will be important for the processes
$h\rightarrow \Upsilon(\bar b b)\,\gamma$,
 $H\rightarrow \Upsilon(\bar b b)\,\gamma$,
where $(m_{b}/m_{\Phi})^{2}\rightarrow$ 0.
 In Figs. 7 and 8 we plot the relative
widths of $h\rightarrow \Upsilon(\bar b b)\,\gamma$ and
$H\rightarrow \Upsilon(\bar b b)\,\gamma$ decays, respectively, compared to
$\bar b b$ final state, vs.  $m_{h}$ and  $m_{H}$, respectively, and
$H\rightarrow T(\bar t t)\,\gamma$ decay (see Fig. 9)
compared to $\bar t t$-state as a function of $m_{H}$.\\
The recognizing of the quark-antiquark bound state can be shown
through the resonance structure having a specific signal, e.g.,
the final leptonic pairs $e^{+} e^{-}$, $\mu^{+}\mu^{-}$ and
 $\tau^{+}\tau^{-}$. This resonance is expected can give one
clear evidence of a bound state production over the QCD
background if the production would be significantly large.

\begin{center}
\begin{figure}[h]
\resizebox{12cm}{!}{\includegraphics{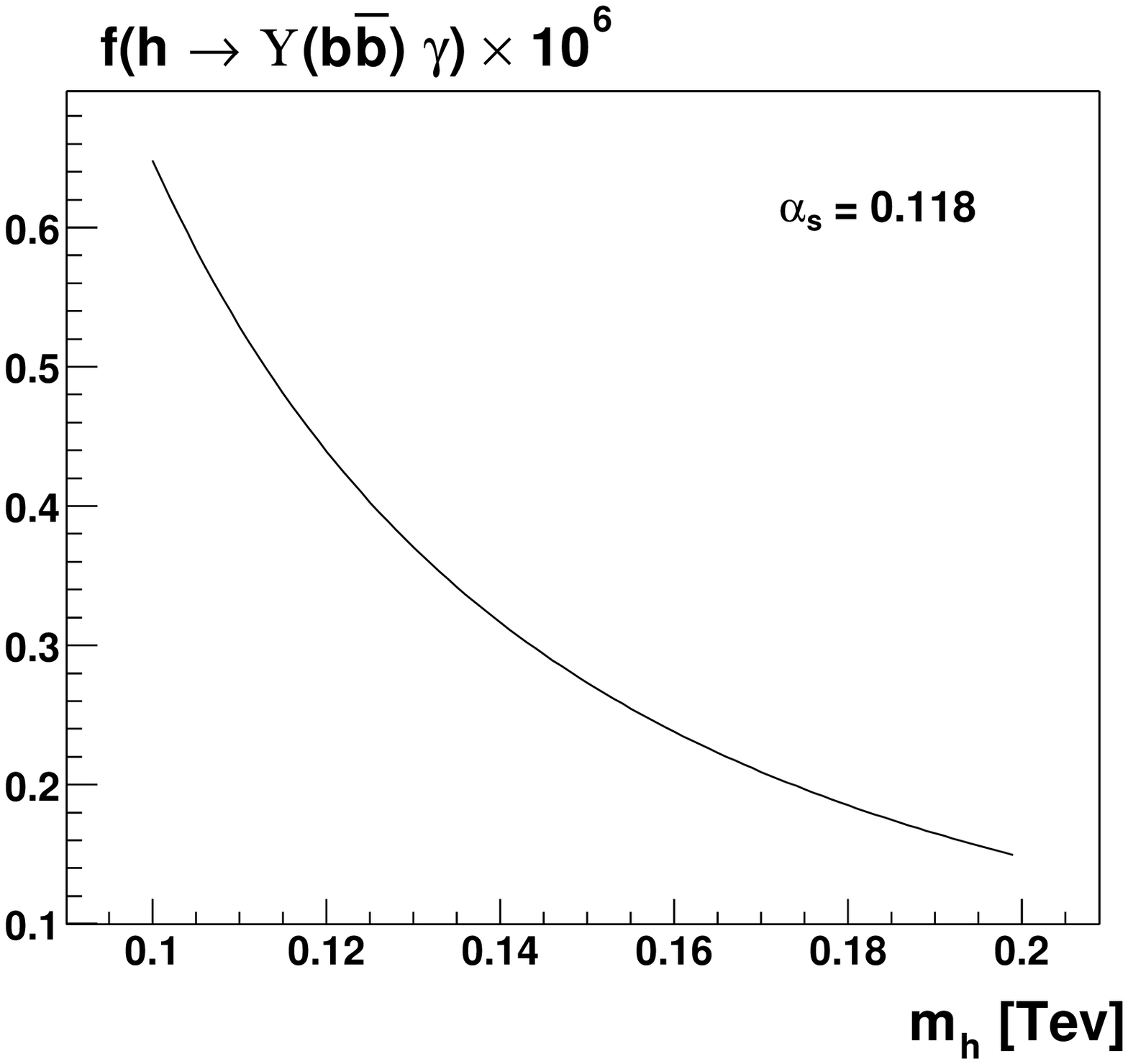}}

Fig.7 The relative decay width $\Gamma (h\rightarrow\Upsilon
(\bar {b} b)\gamma)$ as a function of $h$-boson mass $m_{h}$
compared to $h\rightarrow \bar {b} b$ channel.
\end{figure}
\end{center}

\begin{center}
\begin{figure}[h]
\resizebox{12cm}{!}{\includegraphics{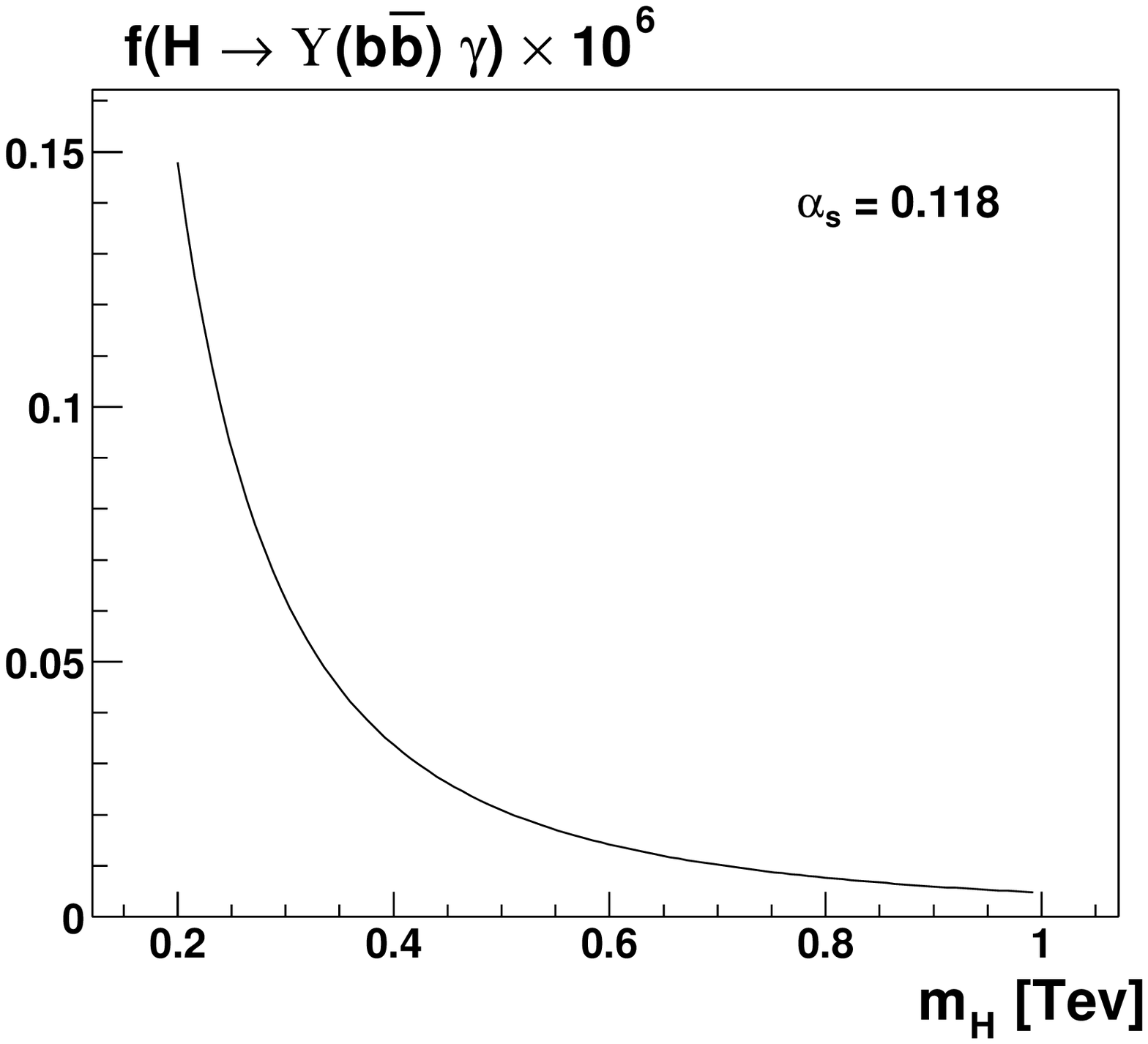}}

Fig.8 The relative decay width $\Gamma (H\rightarrow\Upsilon
(\bar {b} b)\gamma)$ as a function of $H$-boson mass $m_{H}$
compared to $H\rightarrow \bar {b} b$ channel.
\end{figure}
\end{center}

\begin{center}
\begin{figure}[h]
\resizebox{12cm}{!}{\includegraphics{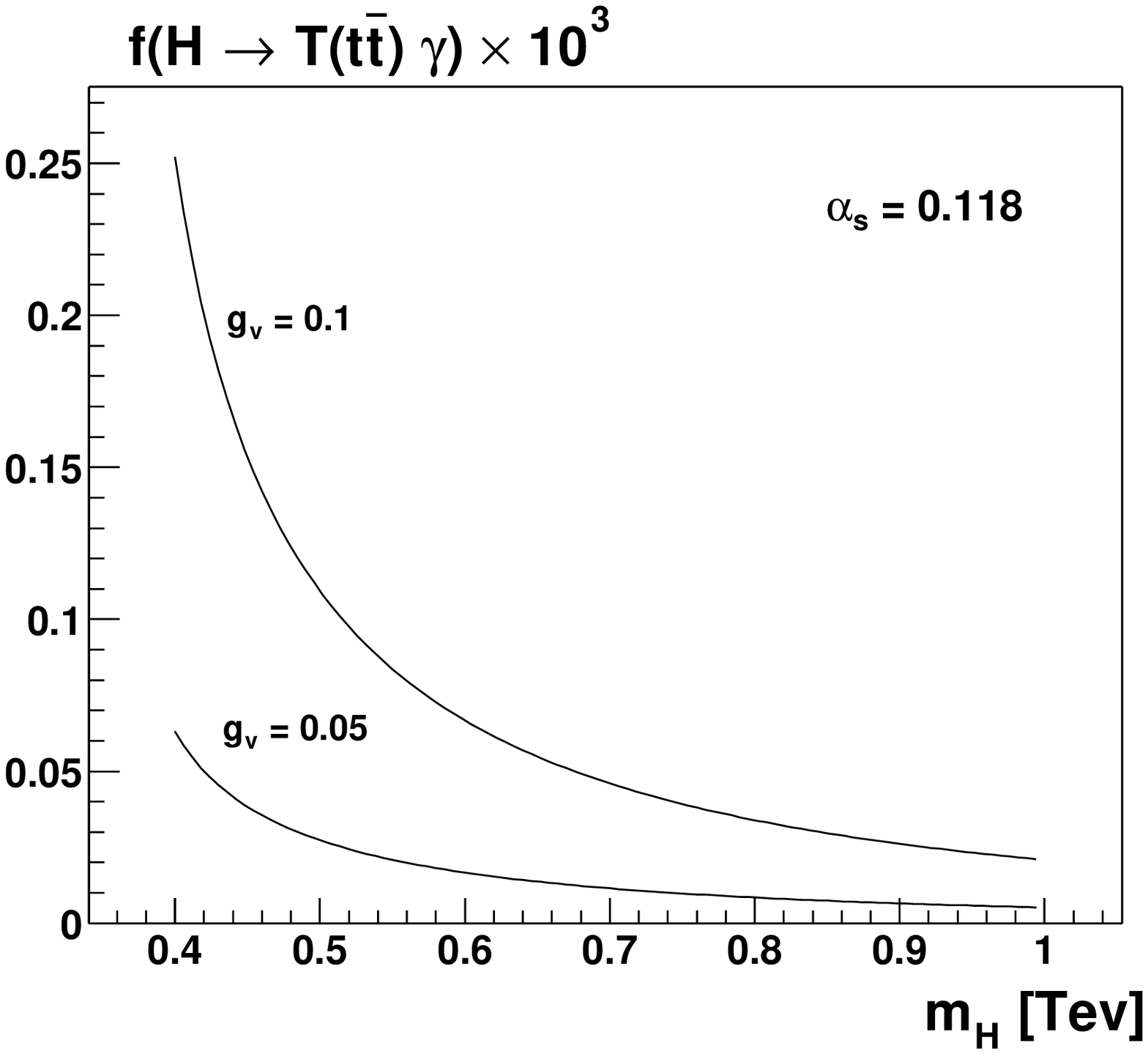}}

Fig.9 The relative decay width $\Gamma (H\rightarrow
T(\bar {t} t)\gamma)$ as a function of $H$-boson mass $m_{H}$
compared to $H\rightarrow \bar {t} t$ channel for different values
of $g_{V}$.
\end{figure}
\end{center}

\section{Effective model for lower-energy theorem}
\noindent
Before proceeding to the lower-energy theorem applied to the transition
$T\rightarrow h Z$ (Model II), we give the general expression for the
amplitude of the decay mentioned above [18]
\begin{eqnarray}
\label{e27}
A(T&\rightarrow& h Z)=\langle\phi_{Z}\vert {\cal L}_{int}
\vert\phi_{T}\rangle\cr
&=&-\frac{1}{v}\left\langle\phi_{Z}\left\vert\sum_{Q=t,U,D}
\frac{\alpha_{s}}{12\,\pi}\,
\eta_{hQ}\,G_{\mu\nu}^{a}\,G^{a\,\mu\nu}-\sum_{q=u,d,s,c,b}\eta_{hq}\,m_{q}\,
\bar q q\right\vert\phi_{T}\right\rangle.
\end{eqnarray}
Here, $\vert\phi_{Z}\rangle$ and $\vert\phi_{T}\rangle$ are
eigenstates responsible
for $Z$-boson and (super)heavy quarkonium $T=T(\bar Q Q)$, respectively.
The second term in
(\ref{e27}) corresponding to the contribution of light quarks comes directly
from the interaction Lagrangian
\begin{eqnarray}
\label{e28}
{\cal L}_{int}=-\frac{h}{v}\left (\sum_{l}\,\eta_{hl}\,m_{l}\,\bar l l+
\sum_{q=u,d,s,c}\,\eta_{hq}\,m_{q}\,\bar q q +
\sum_{Q=b,t,U,D}\,\eta_{hQ}\,m_{Q}\,\bar Q Q- \right.\cr
\left. -2\,\eta_{hW}\,m_{W}^{2}\,W_{\mu}^{+}\,W^{\mu\,-}+
\eta_{hZ}\,m_{Z}^{2}\,Z_{\mu}^{2}\right )\, ,
\end{eqnarray}
while the gluonic contribution given by the operator in the first term
of (\ref{e27}) arises from the coupling of the Higgs boson $h$
to $t$-, $U$- and $D$-quarks through the
standard loop mechanism (see, e.g., references in [18])
with $N_{h}$ number of heavy quarks in the loop. The functions
$\eta_{hi}$ in (\ref{e27}) and (\ref{e28})
($i$=leptons ($l$), light quarks ($q$), $Q$, $W^{\pm}$, $Z$)
are model-dependent ones describing a deviation
from the SM picture where all of $\eta_{hi}$ are equal to unity.
Obviously, considering the fermionic sector, only the terms
containing quarks with masses $m_{Q}\geq m_{h}$ are relevant for giving
rise to the amplitude provided by the matrix element of (\ref{e27})
where the only gluonic part survives.
Our aim is to calculate the amplitude (\ref{e27}). Fortunately, QCD
gives the trace of the energy-momentum tensor $\Theta_{\mu\nu}$ in
the following form [18]:
\begin{eqnarray}
\label{e29}
\Theta_{\mu\mu}=-\frac{b_{6}\,\alpha_{s}}{8\,\pi}\,
G_{\mu\nu}^{a}\,G^{\mu\nu\,a}
+\sum_{q}\,m_{q}\bar q q+\sum_{Q}\,m_{Q}\bar Q Q\, ,
\end{eqnarray}
where $b_{6}$ is the first coefficient of the $\beta(\alpha_{s})$ function
in QCD with 6 flavors of quarks.  At the zero momentum transfer,
the matrix element of (\ref{e29}) between any different eigenstates
$X$ and $Y$ of the Hamiltonian is vanishing [19,18]
\begin{eqnarray}
\label{e30}
\langle X\vert\Theta_{\mu\mu}(q^2)\vert Y\rangle= 0
\end{eqnarray}
with $\langle X\vert\Theta_{00}(q_{0}^2,\vec 0)\vert Y\rangle = 0$. Using the
condition (\ref{e30}), one can obtain the following relation between the
matrix elements containing the gluonic part and the heavy quark terms:
\begin{eqnarray}
\label{e31}
\left\langle\phi_{Z}\left\vert\frac{b_{6}\,\alpha_{s}}{8\,\pi}\,
G^{a}_{\mu\nu}\,
G^{\mu\nu\,a}\right\vert\phi_{T}\right\rangle =
\left\langle\phi_{Z}\left\vert m_{b}\,\bar b b +
m_{t}\,\bar t t\right\vert\phi_{T}\right\rangle \, ,
\end{eqnarray}
where the light quark contributions are neglected. Taking into account
the gluonic anomaly effect through the replacement of the top-quark
mass term [19]
\begin{eqnarray}
\label{e32}
m_{t}\,\bar t t\rightarrow -\frac{2}{3}\frac{\alpha_{s}}{8\,\pi}\,
G_{\mu\nu}^{a}\,
G^{\mu\nu\,a}\, ,
\end{eqnarray}
Eq.(\ref{e31}) transforms into
\begin{eqnarray}
\label{e33}
\left\langle \phi_{Z}\left\vert\frac{b_{5}\,\alpha_{s}}{8\,\pi}\,
G_{\mu\nu}^{a}\,
G^{\mu\nu\,a}\right\vert\phi_{T}\right\rangle =
\left\langle \phi_{Z}\left\vert m_{b}\,
\bar b b\right\vert\phi_{T}\right\rangle\, ,
\end{eqnarray}
where $b_{5}=b_{6}+2/3$ .
Let us consider the amplitude (\ref{e27}) in the following form
\begin{eqnarray}
\label{e34}
A(T\rightarrow h Z)
=-\frac{1}{v}\left\langle\phi_{Z}\left\vert
\sum_{Q=t,U,D}\frac{\alpha_{s}}{12\,\pi}\,
\eta_{hQ}\,G_{\mu\nu}^{a}\,G^{a\,\mu\nu}- m_{b}\,
\bar b b\right\vert\phi_{T}\right\rangle\, ,
\end{eqnarray}
where the terms containing $u,d,s$ and $c$ quarks were omitted
because of its small mass effect on the intermediate quark loop.
Comparing Eqs. (\ref{e34}) and (\ref{e33}), we find the amplitude
for the decay $T\rightarrow h Z$ in the limit of vanishing of
the four-momentum of the Higgs-boson $h$:
\begin{eqnarray}
\label{e35}
A(T\rightarrow h Z)=
-\frac{1}{v}\left\langle\phi_{Z}\left\vert\frac{\alpha_{s}}{8\,\pi}\,
\left (b_{5}-\frac{2}{3}\,\sum_{Q=t,U,D}\,\eta_{hQ}\right )\,G_{\mu\nu}^{a}
\,G^{a\,\mu\nu}\right\vert\phi_{T}\right\rangle\, ,
\end{eqnarray}
where the couplings $\eta_{hQ}$ are [16]:
\begin{eqnarray}
\label{e36}
\eta_{ht(U)}=\frac{\cos\alpha}{\sin\beta}=\sin (\beta-\alpha) + \cot\beta\,
\cos (\beta-\alpha)\, ,
\end{eqnarray}
\begin{eqnarray}
\label{e37}
\eta_{hD}=-\frac{\sin\alpha}{\cos\beta}=\sin (\beta-\alpha) - \tan\beta\,
\cos (\beta-\alpha)\, .
\end{eqnarray}
For numerical estimation, we use the decoupling limit where Eqs. (\ref{e36})
and (\ref{e37}) are transformed into the following ones:
\begin{eqnarray}
\label{e38}
\eta_{ht(U)}\simeq 1 + z\,s_{2\beta}\,c_{2\beta}\,\tan^{-1}\beta\, ,
\end{eqnarray}
\begin{eqnarray}
\label{e39}
\eta_{hD}\simeq 1 - z\,s_{2\beta}\,c_{2\beta}\,\tan\beta\, ,
\end{eqnarray}
where $z=(m_{Z}/m_{A})^{2}$.
To calculate the amplitude (\ref{e35}), one has to estimate its
matrix element as those given by the soft non-perturbative gluonic
field. This contribution occurs as the excitation of the
non-perturbative gluon condensate in an environment of a pair
of a quark and an antiquark bound at the scale
\begin{eqnarray}
\label{e40}
r^{-1}_{bound~ state}\sim\Lambda_{Q}=m_{Q}\,\alpha_{s}
(\tilde\mu\simeq m_{Q}\alpha_{s}),
\end{eqnarray}
which is larger than the scale of strong interactions $\Lambda$
with $\alpha_{s}\ll 1$.   This non-perturbative effect can be
estimated in the transition $T\rightarrow Z$ within a minimal
point-like source $F\,\alpha_{s}\,\vec E^{2}(x)$ [19] with $\vec E$
being the electric component of the gluonic field, and an arbitrary
constant $F$ defines the strength of this source. Hence, the only
remaining work is to calculate the following two-point function
\begin{eqnarray}
\label{e41}
W(G_{\mu\nu})=i\,\int\,d\,x\,e^{i\,q\,x}\left\langle 0\left\vert T\left\{
F\alpha_{s}\,\vec E^{2}(x),\,\frac{\beta(\alpha_{s})}{4\,\alpha_{s}}\,
G_{\mu\nu}^{2}(0)
\right\}\right\vert 0\right\rangle
\end{eqnarray}
embedded in the non-perturbative amplitude $A_{NP}$ of the decay
$T\rightarrow h Z$
\begin{eqnarray}
\label{e42}
A_{NP}(T\rightarrow h Z)=\frac{1}{v\,\Lambda^2}\,
\left (1-\frac{2}{3}\frac{1}{b_{5}}
\sum_{Q=t,U,D}\,\eta_{hQ}\right )\cdot W(G_{\mu\nu})\, .
\end{eqnarray}
Here, $\beta(\alpha_{s})=(-b_{5}\,\alpha_{s}^{2}/2\,\pi)+O(\alpha_{s}^{3})$.
Using the tricks followed by the authors in [12,20], the calculation of the
amplitude (\ref{e42}) can be achieved in the framework of the QCD low-energy
theorem approach in the limit $q^{2}\rightarrow $ 0
\begin{eqnarray}
\label{e43}
A_{NP}(T\rightarrow h Z)=\frac{1}{v}\frac{\pi\,F}{\Lambda^2}\,
\left (1-\frac{2}{3\,b_{5}}
\sum_{Q=t,U,D}\,\eta_{hQ}\right )\,{\left\langle\frac{\alpha_{s}}{\pi}\,
G_{\mu\nu}^{2}\right\rangle }_{0}\, ,
\end{eqnarray}
where ${\langle (\alpha_{s}/\pi)\,G_{\mu\nu}^{2}\rangle }_{0}$ is the standard
gluonic condensate.
Using these formulas, now we can estimate the corrections due to
non-perturbative fluctuations of the gluonic field in the decay
$T(\bar U U)\rightarrow hZ$
\begin{eqnarray}
\label{e44}
\Gamma(T(\bar U U)\rightarrow hZ)=\Gamma _{0}(T(\bar U U)\rightarrow hZ)
\,(1+\delta_{NP})\, ,
\end{eqnarray}
where $\Gamma_{0}(T(\bar U U)\rightarrow hZ)$ is the decay width (\ref{e12}),
while the non-perturbative correction factor $\delta_{NP}$
is defined as $\delta_{NP}=\Gamma_{NP}/\Gamma_{0}$ with
\begin{eqnarray}
\label{e45}
\Gamma_{NP}(T(\bar U U)\rightarrow hZ)=\frac{1}{16\,\pi\,m_{T}}\,
{\vert A_{NP}\vert}^{2}\,
\left (1-\frac{4\,m_{h}^{2}}{m_{T}^{2}}\right )^{1/2}\, .
\end{eqnarray}
The $\delta_{NP}$ correction is obtained at the order of the magnitude of
 $10^{-8}$ for the strength parameter $F=1$.
%
\section{Conclusion and discussion}
%
%
\noindent
In this paper, we have concerned ourselves with the question of existence
of (super)heavy quarkonia and their decays with production of a Higgs-boson,
e.g., $T(\bar U U)\rightarrow hZ$.   The possible existence of the
(super)heavy quarkonium has been studied in the framework of a simple
Higgs-boson potential model (Model I) having a phenomenological "hard"
Yukawa couplings $\lambda (m_{Q},\xi_{\chi\,Q})$ between the (super)heavy
quark and the Higgs-boson $\chi$. Furthermore, to estimate the decay
width of $T(\bar UU)\rightarrow hZ$ by taking into account the
fluctuations of the gluonic field (Model II), we used the conformal
properties of QCD.  We found that the effect of non-perturbative
fluctuations of the gluonic field in the total decay width
$\Gamma(T(\bar U U)\rightarrow hZ)$ is rather small.   $\Gamma_{NP}$ is
quite sensitive to the strength $F$ of the gluonic point-like
source, since it is proportional to square of this strength. The constant $F$
can be estimated within the approach of the Model II through the decay, e.g.,
$T(\bar U U)\rightarrow \bar l l$ $(l=\mu,\tau)$ with
$A_{NP}\sim\langle\bar l l\vert F\,\alpha_{s}\,\vec E^{2}(x)\vert 0\rangle$, if
this decay is known. The amplitude of this process is proportional to $F$,
while it is cancelled in the ratio
$\Gamma(T(\bar U U)\rightarrow hZ)/\Gamma(T(\bar U U)\rightarrow\bar l l)$.
The conformal properties of QCD are especially important in transitions
between lighter hadrons, e.g. $(\bar c c)$- and $(\bar b b)$-bound states,
where the multipole expansion is more sufficient [19].
In the case of (super)heavy quarkonium transitions, the Model I yields an
enhancement effect compared with the result coming from the Model II
at ${\langle(\alpha_{s}/\pi)\,G_{\mu\nu}^{2}\rangle }_{0}
\simeq $ 0.012 $GeV^{4}$.  Such an enhancement of the decay width
$\Gamma (T(\bar U U)\rightarrow hZ)$ with guaranteed
criterion $c=(\Gamma_{tot}/\epsilon_{B}) <1$ could have a quite
significant effect on the shapes of bound-state resonances containing
(super)heavy quarks in the search for new phenomena at the Tevatron and LHC.
Unfortunately, the total decay width $\Gamma_{tot}$ is unknown.  In Fig.4,
we show the dependence of the ratio of its total decay width $\Gamma_{tot}$
to the binding energy $\epsilon_{B}$, i.e. $c=\Gamma_{tot}/\epsilon_{B}$
on the real Yukawa couplings $\xi_{\chi Q}$ in the 2HDM at different
values of $\chi$-boson mass.  We observe that the binding critical
ratio $c$ is strongly sensitive to $\xi_{\chi Q}$. Thus, we believe
that in some kinematical regions there is a possibility for finding
the fourth family quarks at the Tevatron's Run II (the LHC experiments
will not miss it).
If the scale of the Higgs-boson mass does not exceed 180 GeV,
the channels of Higgs decays to $gg$, $\gamma\gamma$ or
$\bar l l$ ($\mu^{+}\mu^{-},\tau^{+}\tau^{-})$ may give an enhancement
effect which can be interpreted as the indication of heavy quark
existence. For example, the decay of the CP-even lightest Higgs-boson
$h\rightarrow gg(\gamma\gamma)$ with $114~GeV<m_{h}<180~ GeV$ could be
enhanced (due to the 4th generation quark contribution) by a factor
$\rho\simeq$ 8.9(4.5-5.5) for $m_{4}$= 200 GeV and $\rho\simeq$
8.5(4.0-5.0) for $m_{4}$= 600 GeV. On the other hand, the decays of
CP-even heavy Higgs $H\rightarrow gg(\gamma\gamma)$ in the mass region
$180~GeV<m_{H}< 800~GeV$ could give $9\leq\rho\leq 13 (15\leq\rho\leq 25)$
for $m_{4}$=200 GeV and $9\leq\rho\leq 27 (15\leq\rho\leq 57)$ for $m_{4}$=
600 GeV. This effect on the $H$ Higgs-boson decay
has a minor dependence of $\tan\beta$. We have shown that a reduction
of a single $U$-quark decay width leads to a significant enhancement
of the signal of the $T(\bar U U)$ resonance.
 We investigated also the possible manifestation of the
 CP-even Higgs-boson states $h$  and $H$ in their rare decays
 $h\rightarrow\Upsilon (\bar b b)\gamma$ and
 $H\rightarrow\Upsilon (\bar b b)\gamma$
or even  the production
of more heavier bound states $T(\bar t t)$ in the rare decays of
Higgs bosons $H$. We obtain that these decays can
be detectable at the forthcoming experiments at the  LHC.

In the final state, the Higgs bosons $h$, the gauge bosons $Z$/$W^{\pm}$
and even charged Higgs-bosons $H^{\pm}$ are on
mass-shell. Hence, the masses of heavy quarks can be reconstructed.
As a trigger, one can choose a semileptonic
decays $U\rightarrow D \bar l\nu_{l}$, $U\rightarrow b \bar l\nu_{l}$,
$t\rightarrow b \bar l\nu_{l}$ ($l=\mu, \tau$) and the leptonic one like
$W\rightarrow l \nu_{l}$ with different lepton flavors to eliminate
the backgrounds $\gamma^{*}, Z, Z^{\prime}\rightarrow \bar l l$.
On the other hand, we suppose that the efficiency for observing
$\bar Q Q$ events
may be high enough because of the characteristic kinematics such as
$U\rightarrow D W$, $U\rightarrow b W$ and $t\rightarrow b W$.
The comparison of the measured decay width mentioned above with
theoretically predicted ones
can exclude or even confirm the fourth family fermions (quarks).
Finally, related to our calculations here, some comments are order
in the following:\\
a. the (super)heavy quarks are considered to be non-relativistic;\\
b. if the $\chi$ Higgs-boson mass is $m_{\chi}\sim {\cal O}(100~GeV)$,
then the decoupling
limit ($\xi_{\chi Q}\sim {\cal O}(1)$) is appropriate only
for $m_{Q}\geq 2\,m_{t}$;\\
c. for the $(\bar t t)$-bound state with $\xi_{\chi Q}\sim O(1)$, the mass
$m_{\chi}$ should be smaller than the lower bound given by the
LEP 2 experiments [21];\\
d. if the heavy quarkonium mass is of the order $2\,m_{t}$, the $\chi$-boson
contribution to the combined potential (\ref{e4}) becomes appreciable only
for large values of $\xi_{\chi t}\geq $ 6.\\
\\
We conclude that the (super)heavy quarkonia effects calculated here can be
significant and should be considered seriously for searching for new
physics beyond the SM.

\section{Acknowledgments}
It is a pleasure to thank Fabiola Gianotti, Nikolai Russakovitch and
 Yulian Boudagov for helpful and stimulating discussions.
We are also acknowledge M. Sher and G. Eilam for informing us of
the works of Refs. [7] and [9].

\end{document}